\documentclass[aps,prb,superscriptaddress,reprint]{revtex4-1}

\usepackage{natbib}
\usepackage{graphicx}
\graphicspath{{figures/}}
\usepackage{epsfig} 
\usepackage{amsmath}

\newcommand\T{\rule{0pt}{2.6ex}}       
\newcommand\B{\rule[-1.2ex]{0pt}{0pt}} 

\usepackage{xcolor}

\begin{document}

\title{Transverse confinement of ultrasound through the Anderson transition in 3D mesoglasses}
\author{L. A. Cobus}
\altaffiliation{Current address: Institut Langevin, CNRS UMR 7587, ESPCI Paris, PSL Research University, 1 rue Jussieu, 75005 Paris, France.}
\affiliation{Department of Physics and Astronomy, University of Manitoba, Winnipeg, Manitoba R3T 2N2, Canada}
\author{W. K. Hildebrand}
\affiliation{Department of Physics and Astronomy, University of Manitoba, Winnipeg, Manitoba R3T 2N2, Canada}
\author{S. E. Skipetrov}
\affiliation{Univ. Grenoble Alpes, CNRS, LPMMC, 38000 Grenoble, France}
\author{B. A. van Tiggelen}
\affiliation{Univ. Grenoble Alpes, CNRS, LPMMC, 38000 Grenoble, France}
\author{J. H. Page}
\affiliation{Department of Physics and Astronomy, University of Manitoba, Winnipeg, Manitoba R3T 2N2, Canada}

\date{\today}

\begin{abstract}

We report an in-depth investigation of the Anderson localization transition for classical waves in three dimensions (3D). Experimentally, we observe clear signatures of Anderson localization by measuring the transverse confinement of transmitted ultrasound through slab-shaped mesoglass samples. We compare our experimental data with predictions of the self-consistent theory of Anderson localization for an open medium with the same geometry as our samples. This model describes the transverse confinement of classical waves as a function of the localization (correlation) length, $\xi$ ($\zeta$), and is fitted to our experimental data to quantify the transverse spreading/confinement of ultrasound all of the way through the transition between diffusion and localization.  Hence we are able to precisely identify the location of the mobility edges at which the Anderson transitions occur.

\end{abstract}

\maketitle

Anderson localization can be described as the inhibition of wave propagation due to strong disorder, resulting in the spatial localization of wavefunctions \cite{Anderson1958,Sheng2006,Abrahams2010}. In the localization regime, waves remain localized inside the medium on a typical length scale given by the localization length $\xi$. Between diffusive and localized regimes there is a true transition, which occurs at the so-called mobility edge and exists only in three dimensions (3D) \cite{Abrahams1979} for systems that respect time reversal and spin rotation symmetry (the so-called orthogonal symmetry class) \cite{EversMerlin2008}. For conventional quantum systems, such as the electronic systems considered in Ref. \cite{Abrahams1979}, this transition to localization occurs when particle energy becomes less than the critical energy. In contrast, the localization of classical waves in 3D is only expected to occur in some intermediate range of frequencies called a {\it mobility gap}: a localization regime bounded by two mobility edges (MEs) (one on either side) \cite{Sheng2006}. This is because localization in 3D requires very strong scattering ($k\ell \sim 1$, where $k$ and $\ell$ are wavevector and scattering mean free path, respectively), and strong scattering is only likely to occur at intermediate frequencies where the wavelength is comparable to the size of the scatterers. Weak scattering, where localization is unlikely, occurs both at low frequencies where the wavelength is large compared to the scatterer size (Rayleigh scattering regime), and at high frequencies where wavelength is small compared with scatterer size or separation (the ray optics or acoustics regime). In the intermediate frequency regime, the scattering strength may vary strongly with frequency due to resonances, and the possibility of localization may be enhanced at frequencies where the density of states is reduced \cite{vanTiggelen1999}. As a result, classical waves may even offer the opportunity to observe many mobility edges (one or more ME pairs) in the same sample.

While searches for Anderson localization in 3D have been carried out for both optical and acoustic waves, acoustic waves offer several important advantages for the experimental observation of localization. Chief among these is the possibility of creating samples which scatter sound strongly enough to enable a localization regime to occur \cite{Hu2008,Hildebrand2014,Aubry2014,Cobus2016b,Hildebrand2015,Cobus2016}. Media which scatter light strongly enough to result in localization have not yet been demonstrated, possibly due to the difficulty of achieving a high enough optical contrast between scatterers and propagation medium \cite{Perspectives2016,Escalante2017}. In addition, effects which can hinder or mask signatures of localization can be bypassed or avoided in acoustic experiments. One of the most significant of these effects is absorption, which has hindered initial attempts to measure localization using light waves \cite{Wiersma1997}. With ultrasound, it is possible to make measurements which are time-, frequency-, and position-resolved, which enable the observation of quantities which are absorption-independent \cite{Page1997}. Inelastic scattering (e.g., fluorescence), which has plagued some optical experiments \cite{Sperling2016}, is also not expected to occur for acoustic waves.

We have reported previously on several aspects of Anderson localization of ultrasound in 3D samples \cite{Hu2008,Faez2009,Hildebrand2014,Aubry2014,Cobus2016b}. In general, we are able to make direct observations of localization by examining how the wave energy spreads with time in transmission through or reflection from a strongly scattering medium \cite{Hu2008,Page2011,Cobus2016b,Cherroret2010}. In this work, we present a detailed experimental investigation of the Anderson transition in 3D, using measurements of transmitted ultrasound. The media studied are 3D `mesoglasses' consisting of small aluminum balls brazed together to form a disordered solid. Results for two representative samples are presented: one thinner and monodisperse, and one thicker and polydisperse. Since we can not perform measurements inside the mesoglass samples, our measurements are made very near the surface. Our experiments measure the {\it transmitted dynamic transverse intensity profile}, which can be used to observe the transverse confinement of ultrasonic waves in our mesoglass samples, and to furthermore prove the existence of Anderson localization in 3D \cite{Hu2008}. By acquiring data as a function of both time and space, this technique enables the observation of (ratio) quantities in which the explicit dependence of absorption on the measurements cancels out, so that absorption cannot obscure localization effects.

Our experimental data are compared with predictions from the self-consistent (SC) theory of localization for open media \cite{Skipetrov2006,Cherroret2010}. This theory is described in Section \ref{theory}, where its development in the context of interpreting experiments such as the ones described in this paper is emphasized. In Section \ref{experimental} we explain the details of our experimental methods for observing the transverse confinement of ultrasound. Section \ref{results} presents new experimental results for two mesoglass samples and their quantitative interpretation based on numerical calculations of the solutions of SC theory for our experimental geometry.  A major focus of this work is to show that this comparison between experiment and theory enables signatures of localization to be unambiguously identified and mobility edges to be precisely located. We aim to provide a sufficiently detailed account of our overall approach that future observations of 3D Anderson localization will be facilitated.

\section{Theory}
\label{theory}
To describe transmitted ultrasound in the localization regime, we use a theoretical model derived from the self-consistent (SC) theory of Anderson localization with a position- and frequency-dependent diffusion coefficient. SC theory was developed by Vollhardt and W\"{o}lfle in the beginning of 1980s \cite{Vollhardt1980,Vollhardt1980a,Vollhardt1992} as a very useful and quantitative way of reformulating the scaling theory of localization \cite{Abrahams1979,Vollhardt1982}. Despite its many successes, the original variant of SC theory had a very approximate way of treating the finite size of a sample $L$ and the boundary conditions at its boundaries. In the return probability, which is the essential ingredient in the SC theory that suppresses diffusion, an upper cut-off $L$ was introduced in the summation over all possible paths in the medium.  This produced the correct scaling of localization with sample size, but is clearly insufficient if one aims at quantitatively accurate results. To circumvent this problem, van Tiggelen \textit{et al.} demonstrated that constructive interference is suppressed by leakage through the boundaries of an open medium, causing the return probability to become position-dependent near the boundaries, and implying the existence of a position-dependent diffusion coefficient \cite{VanTiggelen2000}. The position dependence of $D$ also emerged later from perturbative diagrammatic techniques \cite{Cherroret2008} and the non-linear sigma model \cite{Tian2008}. Subsequent studies focused on the analysis of quantitative accuracy of SC theory in disordered waveguides \cite{Payne2010}, the experimental verification of the position dependence of $D$ \cite{Yamilov2014}, and different ways to improve the accuracy of SC theory deep in the localized regime \cite{Tian2010,Neupane2015}. It should be noted that most of the tests of SC theory with a position-dependent $D$ have been, up to date, performed in 1D or quasi-1D disordered systems, leaving the question about its accuracy in higher-dimensional (e.g., 3D) media largely unexplored.

Here we use self-consistent equations for the intensity Green's function
$C(\mathbf{r}, \mathbf{r}', \Omega) = (4\pi/v_E) \left< G(\mathbf{r}, \mathbf{r}', \omega_0+\Omega/2)
G^*(\mathbf{r}, \mathbf{r}', \omega_0-\Omega/2)\right>$ and the position-dependent diffusion coefficient $D(\mathbf{r}, \Omega)$ derived in Ref.\ \cite{Cherroret2008}:
\begin{eqnarray}
&&\left[-i \Omega - \nabla_{\mathbf{r}} \cdot D(\mathbf{r}, \Omega) \nabla_{\mathbf{r}} \right]
C(\mathbf{r}, \mathbf{r}', \Omega) = \delta(\mathbf{r}-\mathbf{r}'),
\label{sc1}
\\
&&\frac{1}{D(\mathbf{r}, \Omega)} = \frac{1}{D_B} + \frac{12 \pi}{k^2 \ell^*_B} C(\mathbf{r}, \mathbf{r}, \Omega),
\label{sc2}
\end{eqnarray}
where $G(\mathbf{r}, \mathbf{r}', \omega)$ is the Green's function of a disordered Helmholtz equation, $C(\mathbf{r}, \mathbf{r}, \Omega)$ is the return probability, $v_E$ is the energy transport velocity (assumed unaffected by localization effects), $k$ is the wave number, the angular brackets $\left<\cdots\right>$ denote ensemble averaging, and $D_B$ and $\ell^*_B$ are the diffusion coefficient and transport mean free path that would be observed in the system in the absence of localization effects: $D_B = v_E \ell^*_B/3$. As compared to Ref.\ \cite{Cherroret2008}, Eqs.\ (\ref{sc1}) and (\ref{sc2}) are now generalized to allow for anisotropic scattering ($\ell^*_B \ne \ell$) which can be done by repeating the derivation of Ref.\ \cite{Cherroret2008} with $\ell^*_B \ne \ell$ from the very beginning. The result is that $\ell$ is replaced by $\ell^*_B$ in Eq.\ (\ref{sc2}) as follows from the same substitution taking place in the Hikami box calculation in a system with anisotropic scattering \cite{Akkermans2007}.

Physically, the Fourier transform
\begin{eqnarray}
C(\mathbf{r}, \mathbf{r}', t) = \frac{1}{2\pi} \int\limits_{-\infty}^{\infty} d \Omega
C(\mathbf{r}, \mathbf{r}', \Omega) e^{-i \Omega t}
\label{ft1}
\end{eqnarray}
of $C(\mathbf{r}, \mathbf{r}', \Omega)$ gives the probability to find a wave packet at a point $\mathbf{r}$ a time $t$ after emission of a short pulse at $\mathbf{r}'$. The pulse should be, on the one hand, short enough to be well approximated by the Dirac delta function $\delta(t)$ (so that adequate temporal resolution is not sacrificed), but, on the other hand, long enough to ensure the frequency independence of transport properties [such as, e.g., the mean free path $\ell$($\omega$) within its bandwidth]. These two quite restrictive conditions can typically be best fulfilled at long times, when the energy density $C(\mathbf{r}, \mathbf{r}', t)$ becomes insensitive to the duration of the initial pulse.

\subsection{Infinite disordered medium}
\label{TheoryInf}

To set the stage, let us first analyze Eqs.\ (\ref{sc1}) and (\ref{sc2}) in an unbounded 3D medium where $D$ becomes position-independent: $D(\mathbf{r}, \Omega) = D(\Omega)$. The analysis is most conveniently performed in the Fourier space:
\begin{eqnarray}
C(\mathbf{r}, \mathbf{r}', \Omega) &=& \frac{1}{(2\pi)^3} \int d^3 \mathbf{q}
C(q, \Omega) e^{-i \mathbf{q} (\mathbf{r}-\mathbf{r}')}.
\label{ft2}
\end{eqnarray}
Equation (\ref{sc1}) yields
\begin{eqnarray}
C(q, \Omega) &=& \left[ -i \Omega + q^2 D(\Omega) \right]^{-1},
\label{sc1f}
\end{eqnarray}
whereas the return probability $C(\mathbf{r}, \mathbf{r}, \Omega)$ in Eq.\ (\ref{sc2}) is expressed as
\begin{eqnarray}
C(\mathbf{r}, \mathbf{r}, \Omega) &=&
\frac{1}{(2 \pi)^3} \int d^3\mathbf{q} C(q, \Omega)
\nonumber \\
&=& \frac{1}{2 \pi^2} \int_0^{q_{\mathrm{max}}} dq q^2
C(q, \Omega),
\label{sc2f}
\end{eqnarray}
where an upper cutoff $q_{\mathrm{max}}$ is needed to cope with the unphysical divergence due to the breakdown of Eq.\ (\ref{sc1}) at small length scales. The need for the cutoff can be avoided if Eq.\ (\ref{sc1}) is replaced by a more accurate calculation, which indicates that the cutoff is related to the inverse mean free path, a result that is physically intuitive. Equation (\ref{sc1}) is unsatisfactory for length scales $\lesssim \ell^*_B$, leading to the cutoff $q_{\mathrm{max}} = \mu/\ell^*_B$, with $\mu \sim 1$. The precise value of $\mu$ cannot be determined from the present theory, but it fixes the exact location of the mobility edge because one easily finds by combining Eqs.\ (\ref{sc2}), (\ref{sc1f}) and (\ref{sc2f}) that
\begin{eqnarray}
D(0) = D_B \left[1 - \frac{6 \mu}{\pi} \frac{1}{(k \ell^*_B)^2} \right].
\label{d0}
\end{eqnarray}
Hence, a mobility edge (ME) at $k \ell = 1$ (Ioffe-Regel criterion) would correspond to $\mu = (\pi/6)(\ell^*_B/\ell)^2$.

In order to introduce definitions compatible with the experimental geometry of a disordered slab confined between the planes $z = 0$ and $z = L$ of a Cartesian reference frame (see the next subsection for details), the integral in Eq.\ (\ref{sc2f}) can be performed by using a cutoff $q_{\perp}^{\mathrm{max}} = \mu/\ell^*_B$  in the integration over only the transverse component $\mathbf{q}_{\perp}$ of the 3D momentum $\mathbf{q}= \{ \mathbf{q}_{\perp}, q_z \}$:
\begin{eqnarray}
C(\mathbf{r}, \mathbf{r}, \Omega) = \frac{1}{(2\pi)^2} \int\limits_{-\infty}^{\infty} d q_z \int\limits_0^{q_{\perp}^{\mathrm{max}}} d q_{\perp} q_{\perp}
C(q_{\perp}, q_z, \Omega).
\label{ft4}
\end{eqnarray}
This leads to an equation similar to Eq.\ (\ref{d0}):
\begin{eqnarray}
D(0) = D_B \left[1 - \frac{3 \mu}{(k \ell^*_B)^2} \right].
\label{d0perp}
\end{eqnarray}
Now a ME at $k \ell = (k\ell)_c = 1$ would correspond to $\mu = \frac13 (\ell^*_B/\ell)^2$. When fitting the data, we use the link between $\mu$ and ME $(k\ell)_c$ the other way around. Namely, $(k\ell)_c$ will be a free fit parameter to be adjusted to obtain the best fit to the data with $\mu = \frac13 (k\ell)_c^2 (\ell^*_B/\ell)^2$.

In the localized regime $k \ell < (k \ell)_c$, an analytic solution of Eqs.\ (\ref{sc1}) and (\ref{sc2}) can be obtained for a point source emitting a short pulse at $\mathbf{r}' = 0$ and $t' = 0$. To study the long-time limit, we set $D(\Omega) = -i \Omega \xi^2$ and obtain
\begin{eqnarray}
C(\mathbf{r}, \mathbf{r}', t) = \frac{1}{4 \pi \xi^2 |\mathbf{r} - \mathbf{r}'|}
\exp\left(-|\mathbf{r} - \mathbf{r}'|/\xi \right),
\label{longtime}
\end{eqnarray}
where the localization length is
\begin{eqnarray}
\xi = \frac{6 \ell}{(k \ell)_c^2} \left( \frac{\ell}{\ell^*_B} \right)
\frac{\chi^2}{1 - \chi^4}, \;\;\; \chi < 1,
\label{xi}
\end{eqnarray}
and $\chi = k\ell/(k\ell)_c$. When $k$, $\ell$ and $\ell^*_B$ are measured independently or fixed based on some additional considerations, Eq.\ (\ref{xi}) provides a one-to-one correspondence between the value of $(k \ell)_c$ that we obtain from fits to data and $\xi$. It is then convenient to use $\xi$ as the main parameter obtained from a fit to data. In Eq.\ (\ref{xi}), the right-hand side changes sign when the localization transition is crossed  and takes negative values in the diffuse regime $k \ell > (k\ell)_c$. Then Eq.\ (\ref{xi}) can be rewritten as
\begin{eqnarray}
	\zeta = \frac{6 \ell}{(k \ell)_c^2} \left( \frac{\ell}{\ell^*_B} \right)
	\frac{\chi^2}{\chi^4 - 1}, \;\;\; \chi > 1,
	\label{xi_c}
\end{eqnarray}
where $\zeta$ plays the role of a correlation length of fluctuations that develop in the wave intensity when the localization transition is approached.

Equation (\ref{xi}) exhibits one of the problems of SC theory: in the vicinity of the localization transition it predicts $\xi \propto (1 - \chi)^{-1}$ and hence the predicted critical exponent is $\nu = 1$. This value is different from $\nu \simeq 1.57$ established numerically for 3D disordered systems belonging to the orthogonal universality class (see, e.g., Ref.\ \cite{Slevin2014} for a recent review). Recently, the same value of $\nu$ has been found for elastic waves in models that account for their vector character \cite{Skipetrov2018}. To our knowledge, no analytic theory exists that predicts a critical exponent different from $\nu = 1$ \cite{Mirlin2008}.

\subsection{Disordered slab}

To compare theory to experimental data, we need to solve Eqs.\ (\ref{sc1}) and (\ref{sc2}) in a bounded disordered medium having the shape of a slab of thickness $L$, confined between the planes $z=0$ and $z=L$. First of all, Eqs.\ (\ref{sc1}) and (\ref{sc2}) have to be supplemented by a boundary condition corresponding to no incident diffuse flux (since the incident energy is provided by a point source at depth $\ell^*_B$):
\begin{eqnarray}
C(\mathbf{r}, \mathbf{r}', \Omega) - z_0 \frac{D(\mathbf{r}, \Omega)}{D_B}
\left( \mathbf{n} \cdot \boldsymbol{\nabla} \right) C(\mathbf{r}, \mathbf{r}', \Omega) = 0,
\label{bc}
\end{eqnarray}
where $\mathbf{n}$ is a unit inward normal to the surface of the slab at a point $\mathbf{r}$ on one of its surfaces; $\mathbf{n}$ is parallel (antiparallel) to $z$ axis for the surface at $z=0$ ($z=L$). This condition is a generalization of the one derived in Ref.\ \cite{Cherroret2008} to a medium with an arbitrary internal reflection coefficient $R_\text{int}$ that can be obtained using the approach of Ref.\ \cite{Zhu1991}. The so-called extrapolation length $z_0$ is given by
\begin{eqnarray}
z_0 = \frac23 \ell^*_B \frac{1+R_\text{int}}{1-R_\text{int}}.
\label{z0}
\end{eqnarray}

Next, the translational invariance in the $(x, y)$ plane imposes $D(\mathbf{r}, \Omega) = D(z, \Omega)$. We obtain the solution of the system of Eqs.\ (\ref{sc1}), (\ref{sc2}) and (\ref{bc}) in a slab following a sequence of steps described below:
\begin{enumerate}
	
	\item
	Equation (\ref{sc1}) is Fourier transformed in the $(x, y)$ plane:
	\begin{eqnarray}
	C(\mathbf{r}, \mathbf{r}', \Omega) &=& \int\limits_0^{\infty} \frac{d^2 \mathbf{q}_{\perp}}{(2\pi)^2}
	C(q_{\perp}, z, z', \Omega)
	\nonumber \\
	&\times& e^{-i \mathbf{q}_{\perp} (\boldsymbol{\rho}-\boldsymbol{\rho}')},
	\label{ft5}
	\end{eqnarray}
	where $\boldsymbol{\rho} = \{ x, y \}$.
	
	\item
	The resulting equation for $C(q_{\perp}, z, z', \Omega)$, Eq.\ (\ref{sc2}), and the boundary conditions (\ref{bc}) are rewritten in dimensionless variables $\tilde{z} = z/L$, $u = (q_{\perp} L)^2$, $\tilde{\Omega} = \Omega L^2/D_B$, $\tilde{C} = -i \Omega L \times C$, and $d = (D/D_B)/(-i \tilde{\Omega})$:
	\begin{eqnarray}
	\left[1 \right. &+& \left. u d(\tilde{z}, \tilde{\Omega}) \right] \tilde{C}(u, \tilde{z} ,\tilde{z}', \tilde{\Omega})
	\nonumber \\
	\hspace{5mm} &-& \frac{\partial}{\partial \tilde{z}} \left[
	d(\tilde{z}, \tilde{\Omega}) \frac{\partial}{\partial \tilde{z}} \tilde{C}(u, \tilde{z} ,\tilde{z}', \tilde{\Omega}) \right] = \delta(\tilde{z}-\tilde{z}'),
	\label{sc1a}
	\end{eqnarray}
	\begin{eqnarray}
	\frac{1}{d(\tilde{z}, \tilde{\Omega})} &=& - i \Omega
	\nonumber \\
	&+& \frac{3}{(k \ell^*_B)^2} \frac{\ell^*_B}{L}
	\int\limits_0^{u_{\mathrm{max}}} \tilde{C}(u, \tilde{z}, \tilde{z}', \tilde{\Omega}) du,
	\label{sc2a}
	\end{eqnarray}
	\begin{eqnarray}
	\tilde{C}(u, \tilde{z}, \tilde{z}', \tilde{\Omega}) &\pm& i \tilde{\Omega} d(\tilde{z}, \tilde{\Omega}) \tilde{z}_0
	\nonumber \\
	&\times& \frac{\partial}{\partial \tilde{z}} \tilde{C}(u, \tilde{z}, \tilde{z}', \tilde{\Omega})
	= 0,
	\label{bca}
	\end{eqnarray}
	where $u_{\mathrm{max}} = (\mu L/\ell^*_B)^2$ and the signs `+' and `-' in Eq.\ (\ref{bca}) correspond to $\tilde{z} = 0$ and $\tilde{z} = 1$, respectively.
	
	\item
	Equations (\ref{sc1a})--(\ref{bca}) are discretized on grids in $z$ and $u$ (we omit tildes above dimensionless variables from here on to lighten the notation):
	$z_n = (n-1) \Delta z$, with $\Delta z = 1/(N-1)$ and $n = 1, \ldots, N$;
	$u_{\nu} = (\nu-1) \Delta u$, with $\Delta u = u_{\mathrm{max}}/(M-1)$ and $\nu = 1, \ldots, M$:
	\begin{eqnarray}
	&&(\Delta z)^2 \left[ 1 + u_{\nu} d_n(\Omega) \right] C_{nm}(u_{\nu}, \Omega)
	\nonumber \\
	&-& d_n(\Omega) \left[ C_{(n+1)m}(u_{\nu}, \Omega) -
	2C_{nm}(u_{\nu}, \Omega) \right.
	\nonumber \\
	&+& \left. C_{(n-1)m}(u_{\nu}, \Omega) \right]
	- \frac{\Delta z}{2} d'_{n}(\Omega)
	\nonumber \\
	&\times&
	\left[ C_{(n+1)m}(u_{\nu}, \Omega) - C_{(n-1)m}(u_{\nu}, \Omega) \right]
	\nonumber \\
	&=& \Delta z \delta_{nm},
	\label{sc1b}
	\end{eqnarray}
	\begin{eqnarray}
	\frac{1}{d_m(\Omega)} &=& - i \Omega
	\nonumber \\
	&+& \frac{3}{(k \ell^*_B)^2} \frac{\ell^*_B}{L}
	\Delta u \left\{ \sum\limits_{\nu=1}^{M} C_{mm}(u_{\nu}, \Omega)
	\right.
	\nonumber \\
	&-& \left. \frac12 \left[ C_{mm}(u_1, \Omega) + C_{mm}(u_M, \Omega) \right] \right\}
	\label{sc2b}
	\end{eqnarray}
	\begin{eqnarray}
	\Delta z C_{1m}(u_{\nu}, \Omega) &+& i \Omega d_1(\Omega) z_0
	\left[ C_{2m}(u_{\nu}, \Omega) \right.
	\nonumber \\
	&-& \left. C_{1m}(u_{\nu}, \Omega) \right]
	= 0,
	\label{bcb}
	\\
	\Delta z C_{Nm}(u_{\nu}, \Omega) &-& i \Omega d_N(\Omega) z_0
	\left[ C_{Nm}(u_{\nu}, \Omega) \right.
	\nonumber \\
	&-& \left. C_{(N-1)m}(u_{\nu}, \Omega) \right]
	= 0.
	\label{bcc}
	\end{eqnarray}
	Here $d'_{n}(\Omega) = [d_{n+1}(\Omega) - d_{n-1}(\Omega)]/(2 \Delta z)$ for $n = 2, \ldots, N-1$ whereas $d'_{1}(\Omega)$ and $d'_{N}(\Omega)$ are assumed to be equal to $d'_{2}(\Omega)$ and $d'_{N-1}(\Omega)$, respectively.
	
	\item
	We start with an initial guess for $d_n(\Omega)$: $d_n(\Omega) = 1/(-i\Omega)$, corresponding to $D = D_B$. Linear algebraic equations (\ref{sc1b}), (\ref{bcb}) and (\ref{bcc}) are solved for $C_{nm}(u_{\nu}, \Omega)$ at fixed $\Omega$ for all $m = 2, \ldots, N-1$ and $\nu = 1, \ldots, M$. An efficient solution is made possible by the fact that the matrix of coefficients of the system of linear equations (\ref{sc1b}), (\ref{bcb}) and (\ref{bcc}) is tridiagonal; we obtain the solution with the help of a standard routine \verb?zgtsl? from LAPACK library \cite{lapack}. Then, new values for $d_m(\Omega)$ are calculated using Eq.\ (\ref{sc2b}) for $m = 2, \ldots, N-1$. $d_1(\Omega)$ and $d_N(\Omega)$ are found by a linear extrapolation from $d_2(\Omega)$, $d_3(\Omega)$ and $d_{N-2}(\Omega)$, $d_{N-1}(\Omega)$, respectively. In practice, this procedure is performed for $m = 2, \ldots (N+1)/2$ only since $d_m(\Omega)$ is symmetric with respect to the middle of the slab. To increase the accuracy of representation of the integral over $u$ in Eq.\ (\ref{sc2a}) by a discrete sum in Eq.\ (\ref{sc2b}), we use a grid with a variable step $\Delta u$: a small $\Delta u_1 = u_1/(M_1 - 1)$ is used for $u \leq u_1$ and a larger $\Delta u_2 = (u_{\mathrm{max}} - u_1)/(M_2 - 1)$ for $u_1 < u \leq u_{\mathrm{max}}$. The typical values of $u_1$, $M_1$ and $M_2$ used in our calculations are $u_1 = u_{\mathrm{max}}/100$, $M_1 = M_2 = 400$. The number of sites in the spatial grid is typically $N = 2001$. We checked that doubling $M_1$, $M_2$ and $N$ does not modify the results by more than a few percent.
	
	\item
	The solution described in the previous step is repeated iteratively, each new iteration using the values of $d_n(\Omega)$ obtained from the previous one, until either a maximum number of iterations is reached (1500 in our calculations) or a certain criterion of convergence is obeyed (typically, we require that no $d_n(\Omega)$ changes by more than $(10^{-5})$\% from one iteration to another).
	
	\item
	With $d_n(\Omega)$ obtained in the previous step, we solve Eqs.\ (\ref{sc1b}), (\ref{bcb}) and (\ref{bcc}) for the last time for all $\nu = 1, \ldots, M$ and $m = m'$ corresponding to the position $z' = \ell^*_B$ of the physical source describing the incident wave. The corresponding solution $C_{nm'}(u_{\nu}, \Omega)$ allows us to compute the Fourier transforms of position- and time-dependent transmission and reflection coefficients $T(q_{\perp}, \Omega)$ and $R(q_{\perp}, \Omega)$, respectively (we temporarily reintroduce tildes above dimensionless variables for clarity):
	\begin{eqnarray}
	T(q_{\perp}, \Omega) &=& -D(z, \Omega)
	\nonumber \\
	&\times& \left. \frac{\partial}{\partial z} C(q_{\perp}, z, z'=\ell^*_B, \Omega) \right|_{z=L}
	\nonumber \\
	&=& -\frac{\tilde{C}_{Nm'}(u, \tilde{\Omega})}{i \tilde{\Omega} \tilde{z}_0},
	\label{trans}
	\end{eqnarray}
	\begin{eqnarray}
	R(q_{\perp}, \Omega) &=& D(z, \Omega)
	\nonumber \\
	&\times& \left. \frac{\partial}{\partial z} C(q_{\perp}, z, z'=\ell^*_B, \Omega) \right|_{z=0}
	\nonumber \\
	&=& -\frac{\tilde{C}_{1m'}(u, \tilde{\Omega})}{i \tilde{\Omega} \tilde{z}_0},
	\label{refl}
	\end{eqnarray}
	where we made use of boundary conditions (\ref{bc}) to express the derivative of $C$ at a boundary via its value.
\end{enumerate}

The above algorithm allows us to compute $T(q_{\perp}, \Omega)$ and $R(q_{\perp}, \Omega)$ for each $\Omega$. The Fourier transforms of $T(0, \Omega)$ and $R(0, \Omega)$, for example, yield the total time-dependent transmission and reflection coefficients studied in Ref.\ \cite{Skipetrov2006}. The position- and time-dependent intensity in transmission studied in Ref.\ \cite{Hu2008} and used to fit experimental results (Section~\ref{comparison} of this paper) is given by a double Fourier transform
\begin{eqnarray}
T(\boldsymbol{\rho}, t) = \int\limits_{-\infty}^{\infty} \frac{d\Omega}{2\pi}
e^{-i \Omega t}
\int \frac{d^2 \mathbf{q}_{\perp}}{(2\pi)^2} e^{-i \mathbf{q}_{\perp} \boldsymbol{\rho}} T(q_{\perp}, \Omega).\;\;
\label{trans2}
\end{eqnarray}
The dynamic coherent backscattering (CBS) peak $R(\theta, t)$ studied in Ref.\ \cite{Cobus2016b} is obtained more simply as
\begin{eqnarray}
R(\theta, t) &=& R(q_{\perp} = k_0 \sin\theta , t)
\nonumber \\
&=& \int\limits_{-\infty}^{\infty} \frac{d\Omega}{2\pi}
e^{-i \Omega t}
R(q_{\perp} = k_0 \sin\theta, \Omega).
\label{refl2}
\end{eqnarray}

As a final, quite technical, but important remark, we describe our way of performing integrations over $\Omega$ in Eqs.\ (\ref{trans2}) and (\ref{refl2}). These integrations can, of course, be performed by directly approximating integrals by sums and computing $T(q_{\perp}, \Omega)$ and $R(q_{\perp}, \Omega)$ on a sufficiently fine and extended grid of $\Omega$. However, this task turns out to be quite tedious because $T$ and $R$ are oscillating functions of $\Omega$ that decay very slowly as $|\Omega|$ increases. An accurate numerical integration then requires both using a small step in $\Omega$ and exploring a wide range of $\Omega$, which is resource consuming. In order to circumvent this difficulty, we close the path of integration over $\Omega$ in the lower half of the complex plane and apply Cauchy's theorem by noticing that $T(q_{\perp}, \Omega)$ and $R(q_{\perp}, \Omega)$ have special points (poles or branch cuts) only on the imaginary axis. This allows us to deform the integration path to follow a straight line that is infinitely close to the imaginary axis on the right side of it from $\mathrm{Im} \Omega = 0$ to $\mathrm{Im} \Omega = -\infty$ and then a symmetric line on the opposite side of the imaginary axis from $\mathrm{Im} \Omega = -\infty$ to $\mathrm{Im} \Omega = 0$. For CBS intensity we obtain, for example,
\begin{eqnarray}
R(\theta, t) &=& R(q_{\perp} = k_0 \sin\theta , t) =
\frac{-i}{2\pi} \lim\limits_{\epsilon \to 0^+}\int\limits_{0}^{\infty} d\alpha e^{-\alpha t}
\nonumber \\
&\times& \left[ R(q_{\perp} = k_0 \sin\theta, \Omega = -i \alpha + \epsilon)
\right.
\nonumber \\
&-& \left. R(q_{\perp} = k_0 \sin\theta, \Omega = -i \alpha - \epsilon) \right],
\label{refl3}
\end{eqnarray}
where we denoted $\mathrm{Re} \Omega = \pm \epsilon$ and $\mathrm{Im} \Omega = -\alpha$. A similar expression is obtained for the transmitted intensity $T(\boldsymbol{\rho}, t)$ with an additional Fourier transform with respect to $\mathbf{q}_{\perp}$:
\begin{eqnarray}
T(\boldsymbol{\rho}, t) &=&
\frac{-i}{2\pi} \lim\limits_{\epsilon \to 0^+}\int\limits_{0}^{\infty} d\alpha e^{-\alpha t}
\int \frac{d^2 \mathbf{q}_{\perp}}{(2 \pi)^2} e^{-i \mathbf{q}_{\perp} \boldsymbol{\rho}}
\nonumber \\
&\times& \left[ T(q_{\perp}, \Omega = -i \alpha + \epsilon)
\right.
\nonumber \\
&-& \left. T(q_{\perp}, \Omega = -i \alpha - \epsilon) \right].
\label{trans3}
\end{eqnarray}
In the diffuse regime [$k \ell \gg (k \ell)_c$], $R(q_{\perp} = k_0 \sin\theta, \Omega = -i \alpha \pm \epsilon)$ is equal to a sum of Dirac delta-functions representing the so-called diffusion poles, and Eq.\ (\ref{refl3}) is nothing else than the calculation of the integral in Eq.\ (\ref{refl2}) via the theorem of residues. When Anderson localization effects start to come into play for $k \ell$ approaching $(k\ell)_c$ from above, the diffusion poles widen and develop into branch cuts. And finally, in the localized regime [$k \ell < (k \ell)_c$] the different branch cuts that were associated with different diffusion poles, merge into a single branch cut covering the whole imaginary axis.

The advantage of Eqs.\ (\ref{refl3}) and (\ref{trans3}) with respect to Eq.\ (\ref{trans2}) and (\ref{refl2}) is obvious: the presence of the exponential function $\exp(-\alpha t)$ under the integral limits the effective range of integration to small $\alpha$ for the most interesting regime of long times $t$. This allows for an efficient calculation of the long-time dynamics with a reasonable computational effort.\\

Finally, for convenience with comparing theory to experimental data, the output of the SC calculations are scaled in time in units of the diffusion time, \emph{i.e.}, as $t/\tau_D$. For a slab geometry, the diffusion time is related to the leakage rate of energy from the sample, and thus internal reflections play an important role. The diffusion time is defined as
\begin{equation}
\tau_D\equiv\frac{L_{\text{eff}}^2}{\pi^2D},
\label{difftime}
\end{equation}
where $L_{\text{eff}}\equiv L+2z_0$ is the effective sample thickness.

\label{SCtheory}

\section{Experimental}
\label{experimental}
\subsection{Mesoglass samples}

\begin{figure}[t]
	\centering
	\includegraphics[width=\columnwidth]{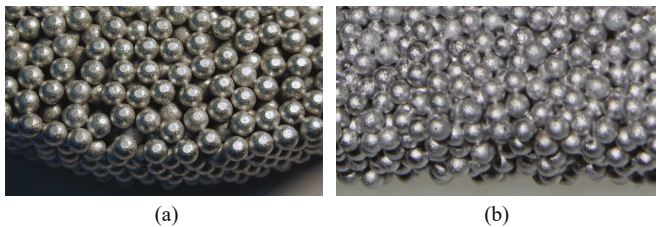}
	\caption{(a) View of a monodisperse mesoglass sample (similar to sample H5). The sample surface has been lightly polished. (b) A polydisperse sample (sample L1). The difference in brazing between the two samples can be seen - the contacts between beads in sample L1 are generally thicker than in sample H5.
		\label{fig1}
	}	

\end{figure}

The `mesoglass' samples examined here are solid disordered networks of spherical aluminum beads, similar to those previously studied \cite{Hu2008,Hildebrand2014,Aubry2014,Cobus2016b,Hu2006,Hildebrand2015,Cobus2016}. These samples are excellent media in which to observe Anderson localization, since the absorption of ultrasound in aluminum is very weak, and the disordered porous structure gives rise to very strong scattering. The samples are slab-shaped, with width much larger than thickness. This geometry is ideal for our measurements; the relatively small thickness enables transmission measurements, while the large width avoids complication due to reflections from the side walls and facilitates the observation of how the wave energy spreads in the {\it transverse} direction, \emph{i.e.}, parallel to the flat, wide faces of the sample. The samples are created using a brazing process which has been described in detail previously \cite{Hu2006,Cobus2016}, resulting in a solid 3D sample in which the individual aluminum beads are joined together by small metal bonds (Fig.~\ref{fig1}). Depending on several factors during the brazing process, the `strength' of the brazing may vary, resulting in thinner/thicker bond joints. This provides a mechanism for controlling the scattering strength in the samples. The entire process is designed to ensure that the spatial distribution of the beads is as disordered as possible \cite{Cobus2016}. In this work, we study two types of brazed aluminum mesoglasses, which differ from each other in terms of bead size distribution and brazing strength, as shown in Fig.~\ref{fig1}. We present experiments and analysis for an illustrative sample of each type: sample H5 is made from  monodisperse aluminum beads (bead diameter is $4.11\pm0.03$~mm), and has a circular slab shape with diameter 120~mm and thickness $L=14.5$~mm. Sample L1 is made from  polydisperse beads (mean bead radius is 3.93~mm with a 20\% polydispersity), is more strongly brazed than sample H5, and has a rectangular slab shape with cross-section $230\times250$~{$\textrm{mm}^2$} and thickness $L=25\pm 2$~mm.

Ultrasound propagates through the samples via both longitudinal and shear components, which become mixed due to the scattering. Our experiments are carried out in large water tanks, with source, sample, and detector immersed in water, and thus only longitudinal waves can travel outside the sample and be detected. As a result, there is significant internal reflection. However, because the waves traveling inside the sample are incident on the boundary over a wide range of angles, the longitudinal ultrasonic waves outside the sample nonetheless include contributions from all polarizations inside the sample. For each experiment, the mesoglass sample is waterproofed, and the air in the pores between beads is evacuated. The sample remains at a low pressure (less than 10\% atmospheric pressure) for the entire duration of the experiment, thus ensuring that the ultrasound propagates only through the elastic network of beads.

To assess the scattering strength in these samples, measurements of the average wave field were performed, allowing results for the phase velocity $v_p$, the group velocity $v_g$ and the scattering mean free path $\ell$ of longitudinal ultrasonic waves to be obtained \cite{Page1996,Cowan1998}.  At intermediate frequencies, our data for $v_p$ and $\ell$ lead to values of $k\ell \sim 1.7$ and $2.7$ for samples H5 and L1, respectively.  These values of $k\ell$ indicate very strong scattering, and are close enough to the Ioffe-Regel criterion $k\ell \sim 1$ to indicate that Anderson localization may be possible in these samples.

\subsection{Time- and position-resolved average intensity measurements}
\label{(MStransmeas)}

\begin{figure}[t]
	\centering
	\includegraphics[width=\columnwidth]{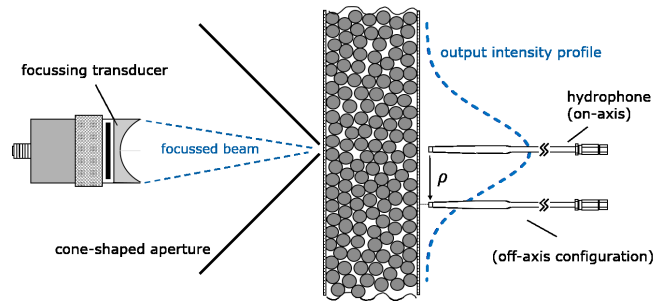}
	\caption{Transverse confinement experimental configuration for a 3D slab mesoglass (the cross-section of which is shown here). A beam is focussed through a small aperture onto the surface of the sample (blue dashed lines, left). The spreading of the output wave energy in the transverse direction (blue dashed line, right) is measured by translating a hydrophone parallel to the sample surface and acquiring the transmitted field near the surface at many transverse positions $\rho$.}
	\label{schematic}
\end{figure}

To investigate the diffusion and localization of ultrasound in our samples, we measure the {\it transmitted dynamic transverse intensity profile}. This quantity is a direct measure of how fast the wave energy from a point source spreads through the sample \cite{Hu2008}. Our experiments measure the transmission of an ultrasonic pulse through the sample as a function of both time and position. The experimental setup is shown in Fig.~\ref{schematic}. On the input side of the sample, a focussing ultrasonic transducer and cone-shaped aperture are used to produce a small point-like source on the sample surface \cite{Page2011b}. Transmission is measured on the opposite side of the sample using a sub-wavelength diameter hydrophone. We denote the transverse position of the hydrophone at the sample surface, relative to the input point, as transverse distance $\rho$. Transmitted field is measured at the {\it on-axis} point directly opposite the source ($\rho=0$), as well as at several {\it off-axis} points ($\rho>0$). From the measured wave field, the time and position dependent intensity, $T(\rho,t)$, is determined (within an unimportant proportionality constant) by taking the square of the envelope of the field. Because time-dependent intensities are measured at all points, they should be affected equally by absorption when compared at the same propagation time. Thus, in the ratio of off-axis to on-axis intensity, absorption cancels \cite{Page1995}. We write this ratio, the normalized transverse intensity profile, as
\begin{equation}
\frac{T(\rho,t)}{T(0,t)}=\textrm{exp}\left(-\frac{\rho^2}{w^2_\rho (t)}\right) ,\\
\label{Ieq}
\end{equation}
where the absorption-independent {\it transverse width}, $w_\rho (t)$, is defined as
\begin{equation}
\frac{w^2_\rho(t)}{L^2}=\frac{-\rho^2/L^2}{\textrm{ln}\left[\frac{T(\rho,t)}{T(0,t)}\right]}.
\label{transwidtheq}
\end{equation}
In the diffuse regime, the transverse intensity profile is Gaussian [see Eq.\ (\ref{Ieq})]; the transverse width is independent of transverse distance $\rho$, and increases linearly with time as $w^2(t)=4D_B t$ \cite{Page1995}. Near the localization regime, however, $w_\rho^2 (t)$ exhibits a slowing down with time due to the renormalization of diffusion, eventually saturating at long times in the localization regime \cite{Hu2008,Cherroret2010}. Close to the localization transition, $w_\rho^2(t)$ depends on $\rho$ (although this dependence is weaker for large $L$), meaning that the transverse intensity profile deviates from a Gaussian shape. It is important to note that this $\rho$-dependence means that the saturation of $w_\rho^2(t)$ in time cannot be simply explained by a time-dependent diffusivity $D(t)$ (which would imply a Gaussian-shaped transverse intensity profile with a $\rho$-independent width), but is a consequence of the position dependence of the diffusion coefficient that is a key feature of Anderson localization in open systems
\cite{Hu2008}.  \\

Because the scattering in our mesoglass samples is so strong, the transmitted signals can be very weak, especially at long times. This means that even very small spurious signals or reflections can influence the data at long times, and it is thus important to ensure that only the signals that were transmitted through the sample are detected by the hydrophone. For each experiment, great care is taken to block any possible stray signals. A cone-shaped aperture (shown in Fig.~\ref{schematic}) is placed at the focal point to block any side lobes from the source spot generated by the focussing transducer. A large baffle, with an opening in its center for the sample, was placed in the water tank between the source and detection side of the sample to block any signals from traveling around the sides of the sample and eventually reaching the detector. Before each experiment, the hole in the baffle was blocked and the hydrophone scanned around the detection side of the tank, to detect any spurious signals from the source; if any were found, their travel path from source to detector was tracked down and blocked. These methods have been described in more detail in Refs.\ \cite{Hildebrand2015,Cobus2016}.

To improve statistics, for each input point, the transmitted field was measured for 4 different $\rho$ values at 13 different $(x,y)$ positions,
\begin{eqnarray}	
(x,y)&=&\{ (0,0),(\pm15,0),(\pm20,0),(\pm25,0),
\nonumber\\
&& (0,\pm15),(0,\pm20),(0,\pm25)\} \textrm{~mm,}
\label{xy}
\end{eqnarray}
where $x$ and $y$ denote transverse positions of the detector in a plane parallel and close to the sample surface (typically a wavelength away), with $\rho=\sqrt{x^2+y^2}$.\\

Our experimental method is designed to facilitate ensemble averaging, which is especially important in the strong scattering or critical regimes where fluctuations play an increasingly important role \cite{Krachmalnicoff2010,Mirlin2000}. Configurational averaging was performed on the data obtained by translating the sample and determining the intensity at all sets of detector positions for each source position. Typically 3025 source positions were recorded for each experiment (a grid of $55\times55$ positions over the sample surface). The source positions were separated by about one wavelength to maximize the number of statistically independent intensity measurements that could be performed on a given sample and ensure that the averaging was not spoiled by spatial correlations \cite{Hildebrand2014}. To reduce the effect of electronic noise, each measurement of the acquired wavefield was repeated many times and averaged together; typically, each signal was averaged 4000--5000 times. As we would like to consider only the multiply scattered signals, any contributions from coherent pulse transport were removed by subtracting the average field from each individual field, \emph{i.e.}, we determine
\begin{eqnarray}
\psi_\text{MS}(t,\rho_\text{in},\rho_\text{out})&=&\psi(t,\rho_\text{in},\rho_\text{out})
\nonumber\\
&&-\left<\psi(t,\rho_\text{in},\rho_\text{out})\right>_{\rho_\text{in}}
\end{eqnarray}\\
and use ${\psi_\text{MS}(t,\rho_\text{in},\rho_\text{out})}$ to obtain the multiply scattered intensities.

\section{Results, analysis and discussion}
\label{results}

\subsection{Amplitude transmission coefficient}
\begin{figure}[t]
	\centering
	\includegraphics[width=\columnwidth]{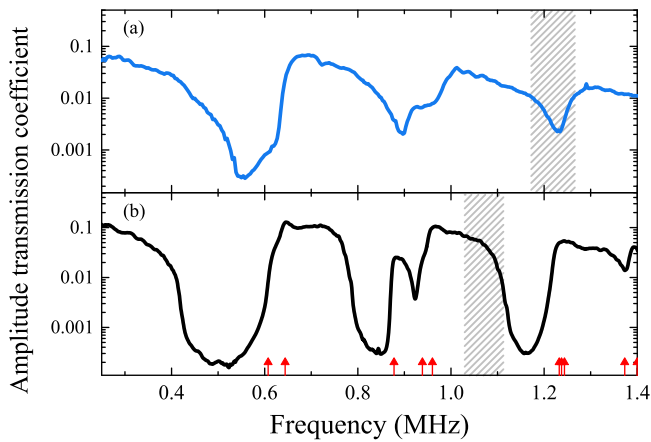}
	\caption{Amplitude transmission coefficient $T_\text{amp}(f)$ as a function of frequency. Data shown are (a) for sample L1, taken using a point-source, and (b) for sample H5, taken using a plane-wave source. Red arrows indicate the resonance frequencies of single, unbrazed 4.11~mm aluminum beads. Vertical grey hatched bars indicate the frequency ranges of interest for transverse confinement analysis.}
	\label{TransCoeff}
\end{figure}

To quantify the frequency-dependence of transmitted ultrasound through our mesoglasses, we calculate the amplitude transmission coefficient $T_\text{amp}(f)$ from the time-dependent transmitted field $\psi(t,\rho=0)$. A Fourier transform converts $\psi(t,\rho=0)$ into the frequency domain, resulting in $\Psi(f)$. The amplitude of $\Psi(f)$ is found, and then configurational averaging is performed on $|\Psi(f)|$ as described in Section~\ref{(MStransmeas)}. The same process, without the configurational average, is performed on the reference field -- the input pulse travelling through water to the detector. The normalized amplitude transmission coefficient is then calculated as:
\begin{equation}
T_\text{amp}(f)=\frac{\left<|\Psi(f)_\textrm{transmitted}|\right>}{|\Psi(f)_\textrm{reference}|} .
\label{tamp}
\end{equation} 
Figure~\ref{TransCoeff}(a) shows $T_{\text{amp}}(f)$ measured this way for sample L1 using a focussed transducer source. It is important to emphasize the difference between this configurational average of the absolute value of the field (in which phase is ignored), and the average field (in which phase coherence plays a significant role, and which gives the effective medium properties).

The amplitude transmission coefficient can also be measured using a plane wave source, approximated by placing the sample in the far field of a flat disk-shaped emitting transducer.  In this case, the transmitted field $\psi(t,x,y)$ is measured with the hydrophone over a large number of positions $(x,y)$ in the speckle pattern [$\sim 11,5000$ positions for the results shown in Fig.~\ref{TransCoeff}(b)], $|\Psi(f,x,y)|$ is averaged over all positions $(x,y)$, and the normalized amplitude transmission coefficient $T_\text{amp}(f)$ is calculated using Eq.\ (\ref{tamp}) [Fig.~\ref{TransCoeff}(b)]. Note that although the overall amplitude of $T_\text{amp}(f)$ changes depending on whether the input is a plane-wave or a point source (due to the normalization of $T_\text{amp}(f)$ which does not account for the finite lateral width of the input beams), the frequency-dependence of $T_\text{amp}(f)$, which is the desired quantity for guiding the interpretation of the experimental results, does not depend on the source used.

In Fig.~\ref{TransCoeff}, the resonance frequencies of single, unbrazed 4.11~mm aluminum beads are shown with red arrows. At long wavelengths ($\lambda\gg d$, where $d$ is the bead diameter), the beads move as a whole, and one might expect the vibrational characteristics to be described by a Debye model with effective medium parameters~\cite{Page2004}. By analogy with a mass-spring system, the beads act as the masses, and small `necks' connecting them act as the springs. The first dip in transmission around 500~kHz corresponds to the upper cutoff frequency for these vibrational modes, which consist only of translations and rotations of the beads. Above the upper cutoff for this long-wavelength regime, when the wavelength becomes comparable with the bead diameter, internal resonances of the beads can be excited, and these bead resonances couple together to form pass bands near and above the individual bead resonant frequencies (Fig.~\ref{TransCoeff}). These pass bands are thus elastic-wave analogues of the "tight-binding" regime for electrons; in the electronic case, tight-binding models of Anderson localization have been extensively used, starting with Anderson's initial paper \cite{Anderson1958}. The width of each pass band is finite since the pass bands do not overlap when the coupling between the beads (determined by the strength of the `necks' between them) is weak. These coupled resonances are the only mechanism through which ultrasound can propagate through the mesoglass in this part of the intermediate frequency regime. Correspondingly, the substantial dips in transmission seen in Fig.~\ref{TransCoeff} are due to the absence of such coupled resonances, and are not related to Bragg effects which would only be expected in media with long-range order, which is not present here. For sample L1, the presence of smaller bead sizes has shifted the transmission dips in $T_\text{amp}(f)$ to higher frequencies, and has lessened their depth compared to the monodisperse sample H5 \cite{Hu2008,Turner1998,Hildebrand2014,Cobus2016b}. These `pseudo-gaps' for L1 are probably also shallower due to slightly stronger brazing between individual beads (Fig. \ref{fig1}) \cite{Lee2014}. In this work, we focus our investigation of Anderson localization on the behaviour at frequencies near the transmission dips seen in the two samples around 1.2 MHz, as indicated by the grey hatched bars in Fig.~\ref{TransCoeff}.

\subsection{Time- position- and frequency-resolved average intensity}
\label{ResultsMS}

\subsubsection{Frequency filtering}
To differentiate precisely between the diffuse, critical, and localized regimes, it is desirable to examine the behaviour of the dynamic transverse profile as the frequency is changed in very small increments. Frequency-dependent results were obtained by first digitally filtering the measured wave fields over a narrow frequency band, by taking the fast Fourier transform of ${\psi_\text{MS}(t,\rho_\text{in},\rho_\text{out})}$, multiplying the resulting (frequency-domain) signal by a Gaussian of the form
\begin{equation}
\label{filteq}
\exp{\left[-(f-f_0)^2/w_f^2\right]}\ ,
\end{equation}
where  $f_0$ is the central frequency of the filter and $w_f$ is the width, and calculating the inverse Fourier transform of the resulting product. By varying the central frequency of the Gaussian window, intensity profiles can then be determined for each frequency.  The width $w_f$ was chosen with the goal of performing sufficiently narrow frequency filtering to resolve the change in behaviour with frequency, without broadening the time-dependent features too much. For the calculation of $T(\rho,t)$ , a typical width of $w_f\sim 15$~kHz was used.

\begin{figure}[t]
	\centering
	\includegraphics[width=1\columnwidth]{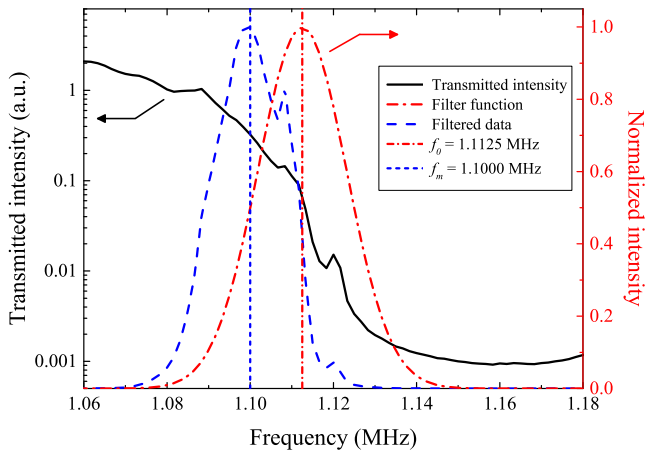}
	\caption{The frequency-pulling effect on a bandwidth-limited signal caused by the frequency-dependence of the average transmitted intensity. The average transmitted intensity (black line) is calculated similarly to the amplitude transmission coefficient of Fig.~\ref{TransCoeff}, but from the average intensity instead of amplitude. The mean frequency of the filtered data $f_m$ is shifted from the central (nominal) frequency of the filter function $f_0$.}
	\label{freqpulling}
\end{figure}

Because the average transmitted intensity varies greatly with frequency, the impact of this dependence on the frequency filtering procedure needs to be assessed. This effect is illustrated in Fig.~\ref{freqpulling}, which shows that, after having been filtered in frequency, the data may not be centered on $f_0$, the nominal central frequency of the filter. In other words, this ``frequency-pulling'' effect means that when the filter function of Eq.\ (\ref{filteq}) is applied to a region where intensity changes rapidly with frequency, the resulting quantity, $T(x,y,t)$, is heavily weighted by data to one side of the central frequency. To account for this shift, the frequency-dependent transmitted intensity is multiplied by the filter function, and the mean frequency of the filtered data $f_m$ is calculated from the first moment of this product. The mean frequency $f_m$ is used to label each set of frequency-filtered data instead of $f_0$, which may not accurately represent the frequency content of the data.

After frequency filtering, the procedure to determine the time-dependent intensity $T(x,y,t)$ is the same as indicated above, namely $T(x,y,t)$ is found by taking the square of the envelope of the time-dependent wave fields. Then, ensemble averaging is performed by averaging the filtered intensity over all $N = 3025$ source positions. The standard deviation in this average is also calculated and divided by $\sqrt{N}$ to give an estimate of the experimental uncertainty in the mean intensity \cite{errorinmean}. The transmitted intensity profiles measured at the same transverse distance from the source position $\rho=\sqrt{x^2+y^2}$ [Eq.\ (\ref{xy})] are averaged together, resulting in average intensity profiles $T(\rho,t)$. Finally, the noise contribution to each averaged $T(\rho,t)$ is estimated from the intensity level of the pre-trigger part of the signal (the signal recorded before the input pulse arrives at the sample input surface, \emph{i.e.}, for $t<0$.). This noise level is subtracted from the average time-dependent intensity, and $w_\rho^2 (t)$ is then calculated using Eq.~(\ref{transwidtheq}).

\subsubsection{Transverse confinement data}
\begin{figure}[t]
	\centering
	\includegraphics[width=\columnwidth]{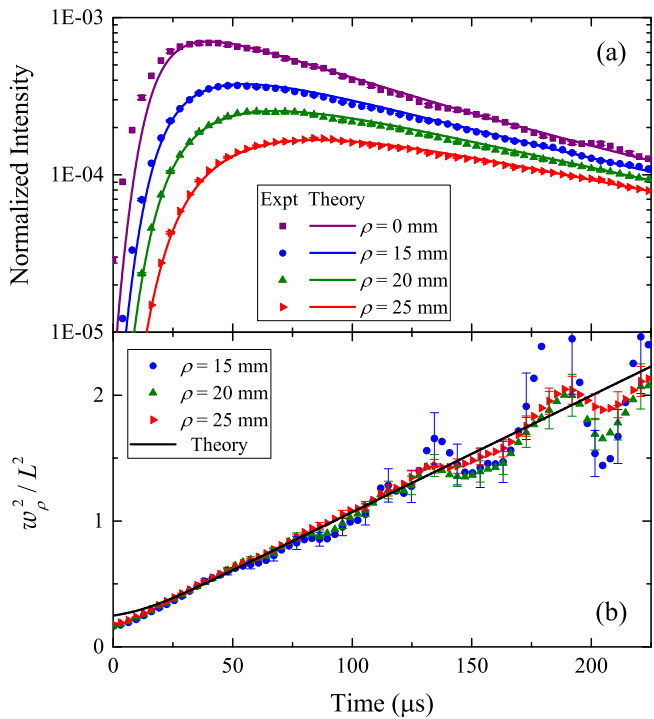}
	\caption{Experimental data (symbols) and fits with diffusion theory (solid lines), for sample L1 at $f_m=250$~kHz. The time-dependence of the transmitted intensity is shown in (a). The time-dependence of $w_\rho^2 (t)$ is shown in (b). The fitting of $w_\rho^2 (t)$ with diffusion theory gives a measure of the Boltzmann diffusion coefficient $D_B=1.45\pm 0.02$~$\textrm{mm}^2/\mu\textrm{s}$. This value of $D_B$ also allows good fits to the time-of-flight profiles $T(\rho,t)$ to be obtained, as shown in (a).   These fits to $T(\rho,t)$ give estimates of the transport mean free path $\ell^*_B\approx 8$~mm and absorption time $\tau_A\approx 560$~$\mu$s. For clarity, error bars are only shown for every 3rd data point.}
	\label{diffusionTOFw2_L1}
\end{figure}

The spreading of wave energy in the sample is characterized by the time- and position-dependent transverse width $w^2_\rho (t)$ [see Eqs.\ (\ref{Ieq}) and (\ref{transwidtheq})]. In the diffuse regime, our experimentally measured $w^2_\rho (t)$ and transmitted intensity profiles $T(\rho,t)$ are well-described by predictions from the diffusion approximation, and may be fit with diffusion theory to ascertain parameters such as $D_B$ \cite{Page1995,Page1997,Cobus2016}. An example of such fitting is shown in Fig.~\ref{diffusionTOFw2_L1}, where data at the low frequency of 250~kHz are reported for sample L1.  The linear time dependence of the width squared, and the observation that the width squared is independent of transverse distance $\rho$ both clearly indicate that the transport behaviour at low frequencies in this sample is diffusive. The slight deviation from linearity in $w^2_\rho (t)$ at early times is due to the finite bandwidth of these frequency-filtered data (35~kHz), as well as to the finite area of the source and detection spots. These finite spot sizes also have the effect of adding a small constant offset to $w^2_\rho (t)$. As emphasized in Ref. \cite{Page1995}, such a measurement of the transverse width provides a direct measurement of the Boltzmann diffusion coefficient $D_B$  without complications due to absorption and boundary reflections. The excellent fit of diffusion theory to the experimental time-of-flight intensity profile $T(\rho,t)$ yields additional information about the transport mean free path and the absorption time \cite{Page1995}.

\begin{figure*}[t]
	\centering
	\includegraphics[width=\textwidth]{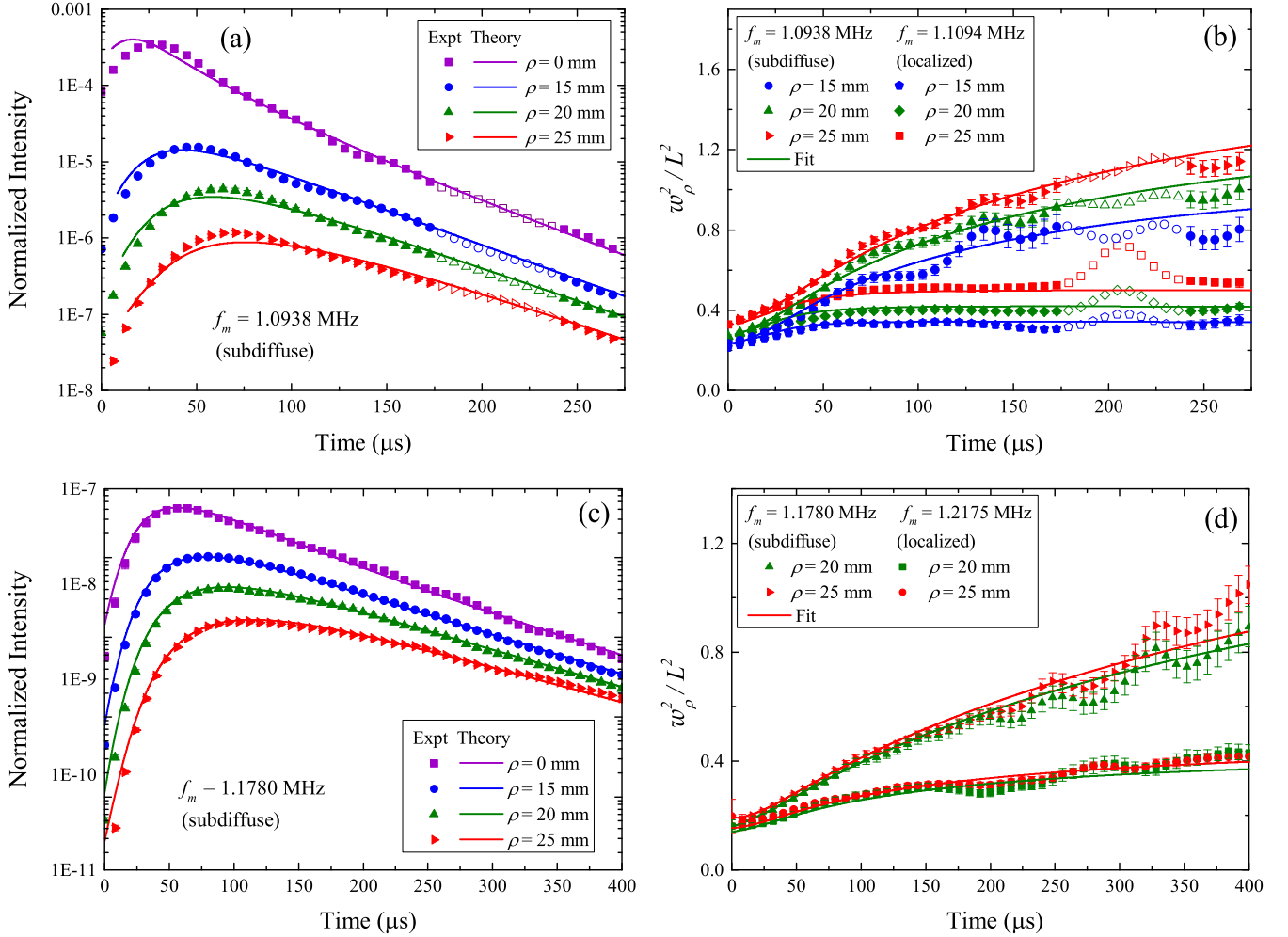}
	\caption{Experimental data (symbols) and fits with SC theory (solid lines), for sample H5 (a,b) and sample L1 (c,d). For some data points, the error bars are smaller than the symbols. Sample H5: The time-dependence of the transmitted intensity at $f_m=1.0938$ MHz is shown in (a). The time-dependence of $w_\rho^2 (t)$ at $f_m=1.0938$ MHz (upper 3 curves) and at $f_m = 1.1094$ MHz (lower 3 curves) is shown in (b). For $f_m=1.0938$ MHz, the best fit result was for correlation length $\zeta=3.63$ cm (diffuse regime). For  $f_m = 1.1094$ MHz, the best fit result was for localization length $\xi = 5.27$ cm (localization regime). In the localization regime, $w_\rho^2 (t)$ increases more slowly and saturates at long times, compared to the diffuse regime. The open symbols show data points that were not included in the fits, due to a measurement artefact that is visible in the data at $f_m=1.1094$ MHz and is discussed in Appendix~\ref{fittingappendix}.  For clarity, only every 80th data point is shown for data in (a,b). Sample L1: The time-dependence of the transmitted intensity at $f_m=1.1780$ MHz is shown in (c). The time-dependence of $w_\rho^2 (t)$ at $f_m=1.1780$ MHz (upper 2 curves) and at $f_m=1.2175$ MHz (lower 2 curves) is shown in (d). Data for only two (of three) $\rho$ values are shown, as the $w_\rho^2 (t)$ curves for different $\rho$ values essentially overlap in this figure (the large sample thickness of L1 substantially weakens the $\rho$-dependence of $w_\rho^2 (t)$; see Eq.~\ref{transwidtheq} and following discussion). For $f_m=1.1780$ MHz, the best fit result for correlation length is $\zeta=3.851$ cm (diffuse regime). For  $f_m=1.2175$ MHz, the best fit result for localization length is $\xi=7.866$ cm  (localization regime) (d). For clarity, only every 100th data point is shown for data in (c,d).}
	\label{diffrhosHL}
\end{figure*}

At higher frequencies, however, the data deviate from the behaviour predicted by the diffusion approximation: notably, $w^2_\rho (t)$ no longer increases linearly with time, but increases more slowly as time progresses, and neither the width squared nor the associated $T(\rho,t)$ curves can be fit with diffusion theory (c.f. Ref.\ \cite{Hu2008}). In the following, we show the evolution of this behaviour as a function of frequency, which is a control parameter for selecting the disorder strength in a single sample. Typical experimental results are shown for both samples in Fig.~\ref{diffrhosHL} (symbols). At these frequencies there are clear deviations from conventional diffusion, as the spreading of wave energy is slower than would be expected if the behaviour were diffusive, and the intensity may become confined spatially as time increases. For sample H5, $w^2_\rho (t)$ even saturates at long times for some frequencies, implying that the transverse spreading of the intensity has halted altogether and suggesting that Anderson localization may have occurred. To determine whether or not this is the case, and to be able to discriminate between subdiffuse and localized regimes, we fit our data with the self-consistent theory of localization.

\subsubsection{Self-consistent theory calculations}
As described in Section~\ref{SCtheory}, our SC theory gives as output the temporally and spatially dependent transmitted intensity $T(\rho,t)$, from which the associated transverse width $w_\rho^2(t)$ can be directly calculated. These SC theory calculations require a number of input parameters, many of which are fixed, as they have been determined from measurements of the average wave field. These fixed input parameters are $\ell$, $k\ell$, and $R_\text{int}$. For simplicity, we use a representative value for each of these parameters in all SC theory calculations for each sample, as determined by an average value appropriate for the frequency ranges of interest (see the grey hatched bars in Fig.~\ref{TransCoeff}). Table~\ref{resultstable} shows values for these average scattering and transport parameters. Our SC calculations do not depend strongly on the values of $\ell$ or $k\ell$ over the range of experimental values used to determine the averages reported in Table~\ref{resultstable}. The internal reflection coefficient $R_\text{int}$ was estimated using a method based on the work of Refs. \cite{Page1995,Ryzhik1996,Turner1995,Zhu1991,Cobus2017}, and its impact on the data analysis is discussed in Appendix~\ref{fittingappendix}.

In addition to these parameters determined from the average field, Table~\ref{resultstable} also includes values for the parameters $L$, $\ell^*_B$ and $\tau_A$: the sample thickness $L$ was measured with calipers and averaged over several sections of the sample, the Boltzmann transport mean free path $l^*_B$ was estimated from SC theory fitting as described in Refs.~\cite{Hildebrand2015,Cobus2016}, and the values of the absorption time $\tau_A$ result directly from fits of SC theory to the time-of-flight profiles $T(\rho,t)$ at the different frequencies of interest (see Appendix~\ref{fittingappendix}). \\
\begin{table}[hb]
	\setlength{\tabcolsep}{.5em}
	\begin{tabular}{|c|c|c|}
		\hline
		& Sample L1 & Sample H5 \T\B \\
		\hline
		$L$ (mm) & 25 & 14.5 \T\B \\
		\hline
		$v_p$ (mm/$\mu$s) & 2.8 & 2.8 \T\B  \\
		\hline
		$v_g$ (mm/$\mu$s) & 2.7 & 2.9 \T\B  \\
		\hline
		$\ell$ (mm) & 1.1 & 0.76 \T\B  \\
		\hline
		$k\ell$ & 2.7 & 1.7 \T\B  \\
		\hline
		$R_\text{int}$ & 0.67 & 0.67 \T\B  \\
		\hline
		$\ell^*_B$ (mm) & 4 & 6 \T\B  \\
		\hline
		$\tau_A$ ($\mu$s) & 170--900 & 100--300 \T\B  \\
		\hline		
	\end{tabular}
	\caption{Acoustic parameters for mesoglass samples in the frequency ranges delineated by the grey hatched bars in Fig.~\ref{TransCoeff}.}
	\label{resultstable}
\end{table}

The final and most important parameter that must be specified to calculate $T(\rho,t)$ and $w_\rho^2(t)$ using the SC theory is $L/\xi$ (or $L/\zeta$).  As indicated in sections~\ref{TheoryInf} and \ref{comparison} this parameter determines how close the predicted behaviour is to the localization transition, where $L/\xi = L/\zeta =0$. The fitting procedure to determine this parameter for a given sample at a given frequency is described in the next section.

\subsubsection{Comparison of data with self-consistent theory}
\label{comparison}

The goal in comparing our experimental data with theory is the determination of the localization (correlation) length $\xi$ ($\zeta$) as a function of frequency. This is achieved by fitting each set of frequency-filtered data with many sets of SC theory predictions, each calculated for a different $\xi$ or $\zeta$ value.  The best fit is found by minimizing the reduced chi-squared $\chi_\textrm{red}^2$. In this way, each set of data, denoted by its unique central frequency $f_m$, is associated with the theory set that fits it best, denoted by its unique value of $\xi$ ($\zeta$). This process is described in detail in Appendix \ref{fittingappendix}.

\begin{figure}[b]
	\centering
	\includegraphics[width=\columnwidth]{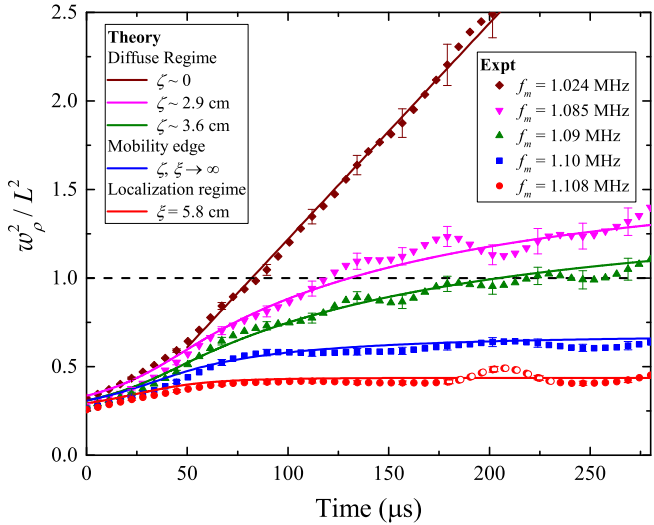} 
	\caption{The time-evolution of $w_\rho^2 (t)$ for sample H5, for one transverse distance $\rho=20$~mm. Fits of self-consistent theory (solid lines) are shown with the data (symbols). Results for five representative frequencies are shown. At 1.024 MHz (brown diamonds), transport is almost entirely diffusive; the slope of the linear fit gives $D = 0.64$~mm$^2$/$\mu$s. As frequency is increased, subdiffuse behaviour is observed (magenta downward pointing triangles, green upward triangles), one then arrives at the mobility edge (blue squares) and finally the localization regime is reached (red circles). Best-fit results from fitting the data with SC theory give the values of the correlation (localization) lengths (legend).}
	\label{w2H}
\end{figure}

Figure~\ref{diffrhosHL} shows representative fitting results for a few frequencies. The predictions of the theory set that best fits the data are shown by the solid lines. For both samples, H5 (top plots) and L1 (bottom plots), the data are well fit by the theory at all times. (Note that the $w_\rho^2 (t)$ curves do not reach zero at $t=0$ due to the effect of the narrow frequency filter width.)

\begin{figure}[h]
	\centering
	\includegraphics[width=\columnwidth]{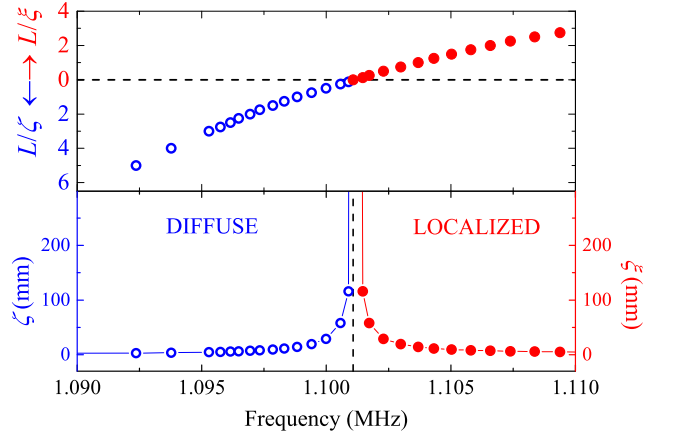}
	\caption{Results from the comparison of self-consistent theory to data, for sample H5 near the mobility edge. The position of the mobility edge is identified as the frequency $f_m$ for which the data is best fit by the SC theory for $L/\xi=0$ (dotted lines). Top plot: the ratio of sample thickness to localization (correlation) length, with $L/\xi(f_m)$ represented by red solid symbols and $L/\zeta(f_m)$ by blue open symbols. Bottom plot: the localization (correlation) length $\xi(f_m)$ (red solid symbols) and $\zeta(f_m)$ (blue open symbols).}
	\label{LxiH}
\end{figure}

\begin{figure}[h]
	\centering
	\includegraphics[width=\columnwidth]{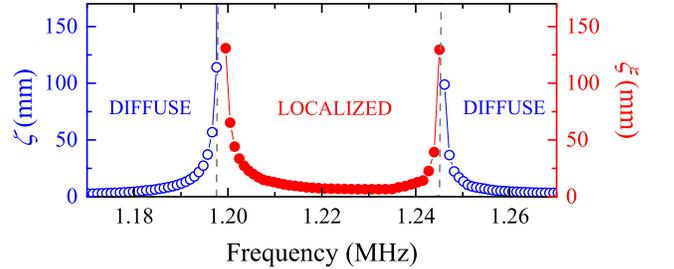}
	\caption{Localization (correlation) length $\xi$ ($\zeta$), as a function of frequency for sample L1. Two mobility edges are identified ($f_m=1.199$ MHz and $f_m=1.243$ MHz), where $\zeta$ and $\xi$ diverge. Between the mobility edges there exists a localization regime.}
	\label{xi_L1}
\end{figure}

Figure~\ref{w2H} shows best theory fits for a single value of $\rho$ at several different frequencies. This figure shows the evolution of $w_\rho^2 (t)$ as the frequency is increased, starting from simple diffuse behaviour, where $w^2 (t)$  increases linearly with time, passing through a sub-diffusive regime where $w_\rho^2 (t)$ increases more slowly, reaching the critical frequency at the mobility edge, where $w_\rho^2 (t)$  saturates in the limit as $t\rightarrow\infty$, and finally crossing into the localized regime, where $w_\rho^2 (t)$  saturates at a constant value in the observation time window. Thus, this figure illustrates how $w_\rho^2 (t)$ reveals the differences in wave transport that are encountered as an Anderson transition for classical waves is approached and crossed in a strongly disordered medium, providing clear signatures of whether or not, and when, Anderson localization occurs.  Furthermore, by determining the best-fit value of $\xi$ or $\zeta$ for each frequency, an estimate of  $\xi(f)$ and $\zeta(f)$ can be obtained, and thus the frequency/ies at which the mobility edge occurs ($L/\xi=0$) can be identified. Results for $\xi(f)$ and $\zeta(f)$ for sample H5 are shown in Fig.~\ref{LxiH}, where a mobility edge can be identified at $f_m=1.101$ MHz. For frequencies above $f_m=1.115$ MHz (deep inside the transmission dip), the level of transmitted signal was not sufficiently above the noise for reliable measurements, and thus only one mobility edge could be identified for sample H5. For sample L1, two mobility edges are identifiable at the critical points where $\xi$ diverges, and the frequency range between them is identified as the localization regime (mobility gap with $L/\xi>0$) (Fig.~\ref{xi_L1}). Whereas only one mobility edge could be identified for sample H5, for sample L1 a measurement of $\xi$ all the way through the mobility gap was obtained. This was possible because sample L1 is polydisperse, with stronger bonds between beads, and thus more signal is transmitted through the sample in the transmission dips (see Fig.~\ref{TransCoeff}) than through sample H5.

\subsubsection{Discussion}

Having identified the localization regime and mobility edge(s) for each sample, we can revisit Figs.~\ref{diffrhosHL} and \ref{w2H}. For frequencies just below the mobility edge  but not yet in the localization regime, the clear deviations from conventional diffusive behaviour are seen, indicating {\it sub-diffusion} when the renormalization of the diffusion coefficient due to disorder hampers the transverse spread of waves  but does not block it entirely. As frequency is increased into the localization regime, the increase of $w_\rho^2 (t)$ with time is initially slower, and eventually saturates at long times. Fig.~\ref{w2H} shows $w_\rho^2 (t)$ for five frequencies near the low-frequency edge of the dip in transmission just below 1.2 MHz. At frequencies where the transmission dip becomes deeper, $w_\rho^2 (t)$ approaches saturation at earlier times, and the data are better fit with theoretical predictions for larger $L/\xi$ values. For sample L1, which is thicker, the range of times experimentally available is not long enough to show a clear saturation of $w_\rho^2 (t)$ (as shown in Fig.~\ref{diffrhosHL}). However, since for each frequency, the best-fit value of the theory to the data gives a measure of the localization length $\xi$ (or $\zeta$ if outside the localization regime), we are still able to determine whether the localization scenario is consistent with our data.

In general, it is important to note that the existence of a transmission dip (Fig.~\ref{TransCoeff}), which is linked to a reduction in the number of coupled resonant modes when the coupling between bead resonances is weak, does not necessarily imply the existence of a mobility gap, which is caused by the interplay between interference and disorder. While it is true that the density of states becomes smaller as the upper edge of a pass band is approached \cite{Lee2014}, and that all mobility edges shown in this work do coincide with the edges of a transmission dip, such a reduction in the density of states may make localization ``easier'' to realize but should not be used on its own as an indication of localization. It is also worth noting that the original evidence of Anderson localization of elastic waves in mesoglasses was found at frequencies outside the transmission dips for these samples \cite{Hu2008}.  We also note that over the entire frequency range studied in this work, our estimates of scattering strength $k\ell$ are consistent with the Ioffe-Regel criterion for localization, which is often interpreted as $k\ell \sim 1$. However, the localization regime only exists in a small section of this spectral region. Thus, a careful and thorough comparison of theory and experiment is essential for determining whether signatures of localization are indeed present.

\begin{figure}[b]
	\centering
	\includegraphics[width=\columnwidth]{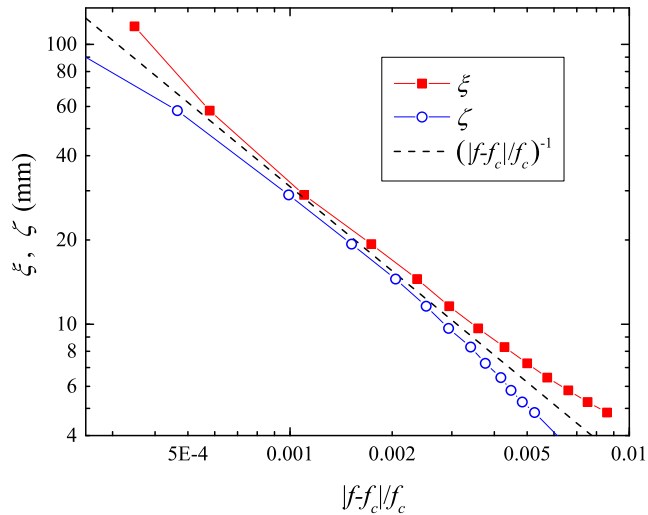}
	\caption{The frequency dependencies of the localization and correlation lengths $\xi$ and $\zeta$ near the mobility edge, from fits of data from sample H5 to the self-consistent theory. Critical frequency $f_{c}$ was found from the fits to be 1.1011 MHz. The power law of $\nu=1$ produced by the self-consistent theory is shown (black dotted line). A power law with $\nu \simeq 0.95$ provides a better fit to the data (not shown). }
	\label{xi_H5}
\end{figure}

Finally, Figs.~\ref{LxiH} and \ref{xi_L1} imply that the critical exponent of the localization transition $\nu$ may be estimated from our results, since near a mobility edge $f_c$, the localization (correlation) length is expected to evolve with $f$ as $\xi(f) \propto |f-f_c |^{-\nu}$. Our measured $\xi$ is shown as a function of $|f-f_c |$ in Fig.~\ref{xi_H5} near the mobility edge at $f_c=1.1011$ MHz for sample H5. The increase of $\xi$ near $f_c$ appears roughly linear, corresponding to a value of $\nu\approx 1$ (shown for comparison in Fig.~\ref{xi_H5} as a dashed line). However, as discussed in Section~\ref{SCtheory}, SC theory itself predicts that $\nu=1$. One might thus argue that that this mean-field value is `built-in', and that therefore our results for the frequency dependence of $\xi$ do not give an independent measurement of the critical exponent. Nonetheless, this outcome (Fig.~\ref{xi_H5}) \textit{does} give additional evidence that our data are consistent with SC theory predictions, and lends additional support to our determination of the locations of mobility edges, which are independent of the exact value of the critical exponent.

\section{Conclusions}

The measurement of the transverse spreading of ultrasound in 3D slab mesoglasses is an excellent method to observe the dynamics of Anderson localization. In particular, in this work we have shown that the width of the transmitted dynamic transverse intensity profile, $w_\rho (t)$, is a sensitive and absorption-independent quantity with which to investigate localization. The transverse width was measured as a function of time and frequency for two different samples. At frequencies approaching the edges of the dips in transmission, we have observed that $w_\rho^2 (t)$ increases less rapidly than linearly with time, tending towards a saturation at long times at frequencies deeper into the transmission dips. This observation agrees with the intuitive expectation that the spreading of wave energy will slow down and eventually halt in the localization regime. We were able to model the slowing of the spread of acoustic energy using the self-consistent theory of localization. Our results show that our experimental measurements agree with the theoretically predicted behaviour for Anderson localization.

The self-consistent theory of localization can provide a detailed quantitative model for our observations. This enabled us to extract several transport parameters of our mesoglass samples. Numerical solutions of the SC theory were obtained and compared to our measurements of the transverse intensity profiles. The comparison of theory and experiment was performed in a careful and systematic way, which enabled us to identify the critical frequency at which the mobility edge occurs, $f_c$. We were able to precisely identify $f_c$ for both samples: for our thinner monodisperse sample, $f_c=1.1011$ MHz while for our thicker, polydisperse sample an entire mobility gap was observed, consisting of a localization regime bounded by two mobility edges at $f_c=1.199$ MHz and $f_c=1.243$ MHz. The comparison of our data with predictions from SC theory is an important strength of this work, as it enabled not only the confirmation of the existence of localization regimes in both samples, but also a complete measurement of the correlation and localization lengths as a function of frequency as the mobility edges were crossed into the localization regimes.

\begin{acknowledgments}
	This work was supported by NSERC (Discovery Grant RGPIN/9037-2001, Canada Government Scholarships, and Michael Smith Foreign Study Supplement), the Canada Foundation for Innovation and the Manitoba Research and Innovation Fund (CFI/MRIF, LOF Project 23523), the Agence Nationale de la Recherche under Grant No. ANR-14-CE26-0032 LOVE, and the CNRS France-Canada PICS project Ultra-ALT.
\end{acknowledgments}

\appendix

\section{Details of the comparison of self-consistent theory with experimental data}
\label{fittingappendix}

In this Appendix, the procedures that were followed to fit predictions of the self consistent theory to the measured transverse widths and time-of-flight profiles are fully described.  To fit one data set with one set of theoretical predictions, we perform a least-squares comparison between experimental $w_\rho^2 (t)$ curves and SC theory predictions. Fits are weighted by the experimental uncertainties. The diffusion time $\tau_D$ [see Eq.~(\ref{difftime})] is a free fit parameter, and is sensitive to the reflection coefficient; however, we have checked that the uncertainty in our estimate of $R_\text{int}$ does not pose a problem for the measurement of $\xi$ or $\zeta$.

To check the reliability in the fitting process, we also fit the intensity profiles $T(\rho,t)$ with theoretical predictions. Thus, each fit is a global fit of both $w_\rho^2 (t)$ and its associated $T(\rho,t)$. Since there are four different $\rho$ values, this yields three $w_\rho^2 (t)$ and four $T(\rho,t)$ curves which are fit simultaneously with the same fit parameters, weighted by experimental uncertainties. Two additional fit parameters are needed only for $T(\rho,t)$; a multiplicative amplitude scaling factor with no physical significance, and the absorption time $\tau_{A}$ which is included by multiplying the theoretical predictions of $T(\rho,t)$ by an additional factor of $\exp(-t/\tau_{A})$ and which, as discussed, cancels out in the ratio used to calculate $w_\rho^2 (t)$ and thus does not affect the $w_\rho^2 (t)$ data.

It is also worth noting several technical but important considerations for the comparison of theory with data. At early times $t$ the self-consistent theory calculations contain known inaccuracies which become worse for larger $\rho$ values. These early times are not included in the fitting procedure, and thus the range of times used for fitting is slightly different for different $\rho$ values. These ranges can be clearly seen in Fig.~\ref{diffrhosHL} (a,b) where the theory curves begin at the earliest times used in the fitting. Late times for which the noise and fluctuations in the data are large are also not included (the latest time in the fits was $275$~$\mu$s for sample H5 and $400$~$\mu$s for sample L1).

Data for sample H5 suffer from an artefact in the acquired signals at some frequencies; just after $200$~$\mu$s, the acoustic signal from the generating transducer has reflected from the front surface of the sample, and travelled back to the generating transducer. This signal induced a small voltage in the piezoelectric generator, which was picked up electromagnetically by the sensitive detection electronics. The narrow-bandwidth frequency filtering applied to the data broadens this (originally brief) signal in time, so a large range of times is affected by this signal. While the artefact is only visible when the signals are small (near the transmission dip), data for this range of times were not included for the fitting at any frequencies for consistency. Data from sample L1 did not suffer from this artefact.

There is a non-negligible effect on the experimental data caused by frequency filtering (Section~\ref{(MStransmeas)}) that must be compensated for in the theory calculations of $T(\rho,t)$.  The filtering operation is equivalent to the convolution of the time-dependent intensity with a function of the form $\textrm{exp}\left[-2(\pi w_f t)^2\right]$, which has the effect of `smearing out' the time-domain signals (see, e.g., early times of  Fig.~\ref{diffrhosHL}). To properly account for this effect, our calculations for $T(\rho,t)$ are convolved with this function before they are used to fit our data.

\subsection*{Estimation of $\xi$ and $\zeta$ from SC theory fitting}

\begin{figure}[b]
	\centering
	\includegraphics[width=0.8\columnwidth]{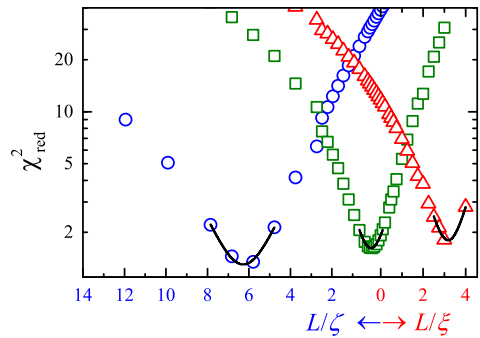}
	\caption{The reduced chi-squared from fitting SC theory to data for three representative frequencies, $\chi_\textrm{red}^2$, is shown as a function of the ratio of sample thickness to localization (correlation) length, $L/\xi$ ($L/\zeta$). For each frequency, the most probable value of $\xi$ or $\zeta$ and its associated uncertainty are found via a parabolic fit near the minimum point [Eqs.~(\ref{chi2}) and (\ref{sigmachi})]. The three representative frequencies are $f_m=1.1780$ MHz (diffuse regime, blue circles), $f_m=1.1968$ MHz (very close to the mobility edge, green squares), and $f_m=1.2175$ MHz (localized regime, red triangles).}
	\label{parabfitting}
\end{figure}

As outlined in Section \ref{comparison}, $\xi(f)$ and $\zeta(f)$ are determined by comparing all sets of frequency-filtered data (each with a unique central frequency $f_m$) with all sets of calculated SC predictions (each with a unique value of $\xi$ or $\zeta$). For each fit, the reduced chi-squared is recorded; the best fit is the one with the smallest  $\chi_\textrm{red}^2$. By filtering the data in frequency with a very fine resolution (for many, closely spaced, central frequencies $f_m$), and calculating many sets of theory over a wide range of closely spaced $\xi$ and $\zeta$ values, it is possible to estimate $\xi(f_m)$ and $\zeta(f_m)$ with precision. However, this method requires a great deal of time-intensive data processing and fitting. A more efficient approach is to estimate the {\it most probable} value of $\xi$ or $\zeta$ for each $f_m$. To do this, we consider the reduced chi-squared  results from the least-squares comparison of data with theory. Figure~\ref{parabfitting} shows $\chi_\textrm{red}^2$ for three different sets of data (each at a different frequency $f_m$) for sample L1. For example, in the localization regime, the best fit, i.e. the most probable value $\xi_\textrm{best}$, corresponds to the minimum of the function
\begin{equation}
	\chi_\textrm{red}^2\propto (\xi-\xi_\textrm{best})^2 /\sigma^2,
	\label{chi2}
\end{equation}
and, for a sufficiently large data set, the uncertainty in the most probable value is given by the curvature of this function near its minimum \cite{Bevington2002}
\begin{equation}
	\sigma^2=2\left(\frac{\partial^2\chi_\textrm{red}^2}{\partial \xi^2}\right)^{-1}.
	\label{sigmachi}
\end{equation}
Similar expressions in terms of $\zeta$ apply in the diffuse regime. This formalism can be applied to our results to estimate the most probable value of $\xi$ or $\zeta$ for each frequency, based on the available data and theory, by fitting a parabola to a few points around the minimum value of $\chi_\textrm{red}^2$ \cite{Bevington2002} (Fig.~\ref{parabfitting}). This method does not require the data to be filtered with closely spaced values of $f_m$, reducing the required calculation time (for the frequency filtering of the data and fitting theory to experiment) and amount of filtered data. In addition, the parabola-fitting technique gives an estimate of the uncertainty for the resulting estimates of $\xi$ ($\zeta$). However, this uncertainty measure most likely underestimates the actual uncertainty in our results for $\xi$ ($\zeta$), as it does not take into account the effects of uncertainties in the estimates of parameters such as $R_\text{int}$ or $k\ell$.


\begin{thebibliography}{53}
\bibitem [{err()}]{errorinmean}%
  \BibitemOpen
  \href@noop {} {}\bibinfo {note} {In the localization regime, it is likely
  that our estimates of uncertainty in the mean intensity are underestimated
  due to the enhanced probability of long-range spatial correlations. We do not
  believe this to affect the validity of our comparison of data with SC theory
  predictions.}\BibitemShut {Stop}%
\end{thebibliography}
\end{document}